\newcommand\FPN{F$_2$PNNNO}
\newcommand\FPNformula{
2-[2$^\prime$,6$^\prime$-difluoro-4$^\prime$-($N$-\emph{tert}-butyl-$N$-oxyamino)phenyl]-4,4,5,5
-tetramethyl-4,5-dihydro-$1H$-imidazol-1-oxyl}

\documentclass[twocolumn,prb,floatfix,preprintnumbers,amsmath,amssymb,superscriptaddress]{revtex4}

\usepackage{graphicx}% Include figure files
\usepackage{dcolumn}% Align table columns on decimal point
\usepackage{bm}% bold math

\begin{document}

\title{Spin density distribution in a partially magnetized organic quantum magnet}

\author{A. Zheludev}
 \email{zheludevai@ornl.gov}
\author{V. O. Garlea}
\affiliation{Neutron Scattering Sciences Division, Oak Ridge
National Laboratory, Oak Ridge, Tennessee 37831-6393, USA.}

\author{S. Nishihara}
\author{Y. Hosokoshi}
\affiliation{Department of Physical Science, Osaka Prefecture
 University, ,Osaka 599-8531, Japan.}
\affiliation{Institute for Nanofabrication Research, Osaka
 Prefecture University, Osaka 599-8531, Japan.}

\author{A. Cousson}
\author{A. Gukasov}
\affiliation{Laboratoire Leon Brillouin, CEA-CNRS Saclay,
 France.}

\author{K. Inoue}
\affiliation{Department of Chemistry, Hiroshima University,
 Hiroshima 739-8526, Japan.}

\date{\today}

%=====================================================================
%=====================================================================
%=====================================================================
%=====================================================================
%=====================================================================

\begin{abstract}
Polarized neutron diffraction experiments on an organic magnetic
material reveal a highly skewed distribution of spin density
within the magnetic molecular unit. The very large magnitude of
the observed effect is due to quantum spin fluctuations. The data
are in quantitative agreement with direct diagonalization results
for a model spin Hamiltonian, and provide insight on the actual
microscopic origin of the relevant exchange interactions.
\end{abstract}

\maketitle

%=====================================================================
%=====================================================================
%=====================================================================
%=====================================================================
%=====================================================================

\section{Introduction}

Static spin correlations in low-dimensional and molecular magnets
are often severely affected by zero-point quantum spin
fluctuations. The simplest and most extreme example is that of an
antiferromagnetic (AF) spin dimer, where the spin density
distribution $\mathbf{S}(\mathbf{r})$ is strictly zero in the
ground state, even in the presence of a small aligning magnetic
field. In partially magnetized states of more complex systems one
can expect to find exotic non-trivial spin densities that too are
strongly modified by quantum fluctuations. Experimental studies of
these correlations can be revealing of the underlying physics, and
help determine or validate theoretical models used to describe
such materials. Below we report a direct  measurement of
field-induced magnetization densities in the novel organic
molecular magnet \FPNformula, \FPN\ for short.

This compound is a prototypical spin-tetramer
system.\cite{Hosokoshi1998,Hosokoshi1999} Its molecular building
block (Fig.~1a) contains only $s$- and $p$- elements, but is
nevertheless magnetic, thanks to two unpaired electrons that
reside in $\pi^\ast$ antibonding molecular orbitals of the
nitronyl nitroxide (NN) and the {\it tert}-butyl nitroxide (tBuNO)
groups, respectively. The material crystallizes in an orthorhombic
structure,  $a=19.86$~\AA, $b=14.01$~\AA\ and $c=13.48$~\AA, space
group P$_{bca}$. In the crystal, \FPN\ molecules are arranged in
pairs, so that their tBuNO groups are close enough for partial
orbital overlap, enabling inter-molecular magnetic interactions.
The result is a two-molecule unit containing four interacting
spins (Fig.~1b). A model Heisenberg Hamiltonian for these
$S=1/2$-tetramers was proposed based on bulk susceptibility and
high field magnetization data.\cite{Hosokoshi1999} Intra-molecular
exchange coupling is ferromagnetic, with $J_\mathrm{F}\approx
35$~meV. Inter-molecular interactions are AF in nature, of
magnitude $J_\mathrm{AF}\approx 5.8$~meV.\cite{Hosokoshi1999} In
agreement with experiment, this model has a unique non-magnetic
ground state with total spin $S_\mathrm{total}=0$, and a gap in
the excitation spectrum.  The magnetization density in the ground
state is strictly zero in entire space.

\begin{figure}
\includegraphics[width=8.7cm]{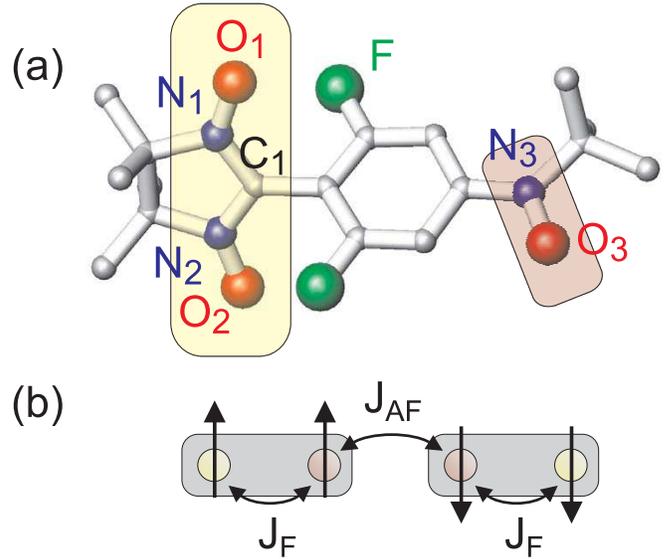}
\caption{(Color online) (a) Molecular structure of \FPN. Hydrogen
atoms are not shown. The shaded rectangles represent
$S=1/2$-carrying unpaired electrons distributed over the nitronyl
nitroxide and {\it tert}-butyl nitroxide groups. (b) A schematic
representation of the 4-spin Heisenberg Hamiltonian for a
two-molecule \FPN\ unit. $J_\mathrm{F}$ and $J_\mathrm{AF}$ are
ferro- and antiferromagnetic exchange interactions, respectively.
Vertical arrows represent individual spins in the classical ground
state $|\uparrow \uparrow \downarrow \downarrow \rangle$. The
actual quantum ground state has zero spin density throughout the
tetramer.} \label{fig1}
\end{figure}

\begin{figure}
\includegraphics[width=8.7cm]{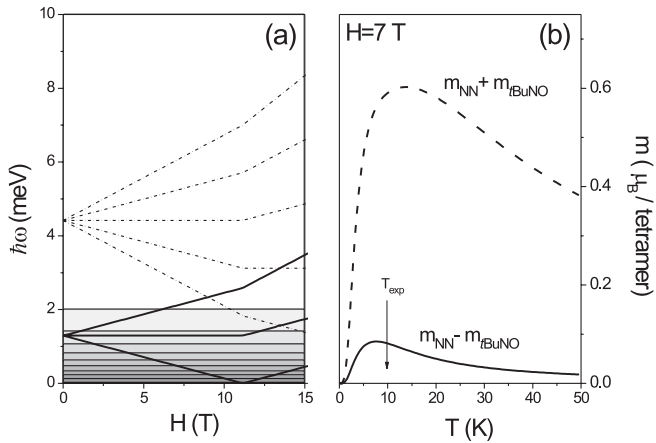}
\caption{ (a) Calculated field-dependence of the energy levels of
an \FPN\ spin-tetramer. Heavy solid lines are the lowest energy
$S=0$ and $S=1$ states. Dashed lines are $S=2$ states. More
excited states are present above 35~meV. The plot is laid over a
filled-contour plot of the thermal population function
$\exp(-\hbar \omega/T)$ for $T=10$~K (0.1 contour step). (b)
Calculated temperature dependence of the sum (dashed line) and
difference (solid line) of the spin populations of the nitronyl
nitroxide and {\it tert}-butyl nitroxide groups. The aroow
indicates the experimental temperature.} \label{tetramer}
\end{figure}

A non-trivial  spin (magnetization) density is only to be found in
some of the tetramer's {\it excited} states. In the presence of an
external magnetic field applied along the $z$ axis, the one with
the lowest energy has a total spin $S_\mathrm{total}=1$ and a spin
projection $S_{z,\mathrm{total}}=+1$. We shall denote this state
as $|1,+1\rangle$. By numerically diagonalizing the 4-spin
Heisenberg Hamiltonian we find that it actually is a linear
combination of four ``pure'' spin wave functions:
\begin{equation}
 |1,+1\rangle= \alpha |\uparrow \uparrow \uparrow \downarrow
 \rangle + \beta |\uparrow \uparrow \downarrow \uparrow
 \rangle - \alpha |\downarrow \uparrow \uparrow \uparrow
 \rangle - \beta |\uparrow \downarrow \uparrow \uparrow
 \rangle, \label{wave}
\end{equation}
where $\alpha \approx 0.46$ and $\beta \approx 0.54$. The most
striking consequence of the quantum-mechanical nature of this
state is a skewed  spin density distribution $S_z(\mathbf{r})$.
The local spin populations of the NN groups are expected to be
equal, but  different from those of the tBuNO groups. Their ratio
$R$ is given by $R=\alpha^2/\beta^2\approx 1.39$. The central
purpose of this work is an experimental detection of this effect.

\section{Experimental}

To observe this phenomenon, one must first prepare the tetramer in
its first excited state. One strategy is to substantially increase
the external field. Due to Zeeman effect, the energy of the
excited state will decrease, and eventually reach zero at some
critical field $H_c$. At this point it will become the {\it new}
ground state, for which the spin density distribution can be
measured. In \FPN, due to residual inter-tetramer interactions,
the transition at $H_c$ is spread out between $H_{1}=9$~T and
$H_{2}=15$~T.\cite{Hosokoshi1999} While it is certainly possible
to perform experiments at $H>H_{1}$, the equipment available for
the present study was limited to fields up to $7$~T. For this
reason we used a slightly different approach. First, a high field
was used to lower the energy of the $|1,+1\rangle$ state as much
as possible. The data were then taken at an elevated temperature
of $T=10$~K that made this state partially populated due to
thermal fluctuations. Of course, states $|1,0\rangle$ and
$|1,-1\rangle$, as well as other higher-spin states got thermally
excited as well. However, at $T=10$~K thermal populations of
higher-spin states are negligible. This is illustrated in
Fig.~\ref{tetramer}(a) that shows the field dependence of tetramer
energy levels calculated using exact numerical diagonalization of
the Heisenberg Hamiltonian. The plot is laid over a shaded contour
plot of the thermal population function $\exp(-\hbar \omega /T)$.
At $H=7$~T all but the lowest $S=1$ energy levels are outside the
populated region. The $|1,0\rangle$ state plays no role, as it has
$S_z(\mathbf{r})\equiv 0$. The spin density distribution in
$|1,-1\rangle$ is exactly the reverse of that for $|1,+1\rangle$,
and does not affect the imbalance between the NN and tBuNO groups.
The only adverse effect of the finite-$T$ approach is a reduction
of the total magnetization of the tetramer, that ultimately
reduces signal strength in any spin density measurement. The
actual value of $T=10$~K was selected to optimize both the total
tetramer magnetization and the predicted population difference
between the NN and tBuNO groups at $H=7$~T (Fig.~\ref{tetramer}b).
A full thermodynamic calculation for the 4-spin Hamiltonian
predicts $R=1.32$ for these conditions, just slightly less than
the ideal value $R=1.39$ for a tetramer purely in the
$|1,+1\rangle$ state.

Measuring the distribution of about 0.5~$\mu_\mathrm{B}$ over two
molecules with 48 atoms each with angstrom resolution is a
formidable experimental challenge. It can only be met by polarized
neutron diffraction.\cite{Gillon1989} This technique achieves
great sensitivity by exploiting the interference between magnetic
and nuclear scattering of neutrons in the crystal. For \FPN\ the
data were taken at the 5C1 and 6T2 lifting counter diffractometers
installed at the Orpheé reactor at LLB, using 0.841~\AA\ and
1.4~\AA \ neutrons, respectively. Beam polarizations of 91\% or
97\% were achieved using Heussler-alloy monochromator and
supermirror bender. A 20~mg \FPN\ single crystal sample was
mounted consecutively with the $a$, $b$ and $c$ axes parallel to
the field direction. Sample environment was a split-coil
cryomagnet. Overall, 70 independent flipping ratios were measured
in magnetic fields $H=7$~T and $H=4$~T at T=10 K, typically
counting 2 hours per reflection on the 5C1 and 30 minutes on the
6T2 diffractometers, respectively. Extracting the corresponding
spatial Fourier components of $S_z(\mathbf{r})$ from these data
required knowledge of the low-temperature crystal structure. The
latter was measured in a single crystal unpolarized neutron
diffraction experiment.  3887 independent Bragg intensities were
measured for a 5 mg single crystal sample on the 5C2 4-circle
diffractometer at Orpheé using 0.832 Å neutrons. Sample
environment was a gas flow cryostat, and the data were taken at
T=50 K. The crystal structure was refined assuming isotropic
vibrational parameters for hydrogen atoms, and general anisotropic
ones for all other. The resulting least-squares R-factor was
0.089.

\begin{figure}
\includegraphics[width=8.7cm]{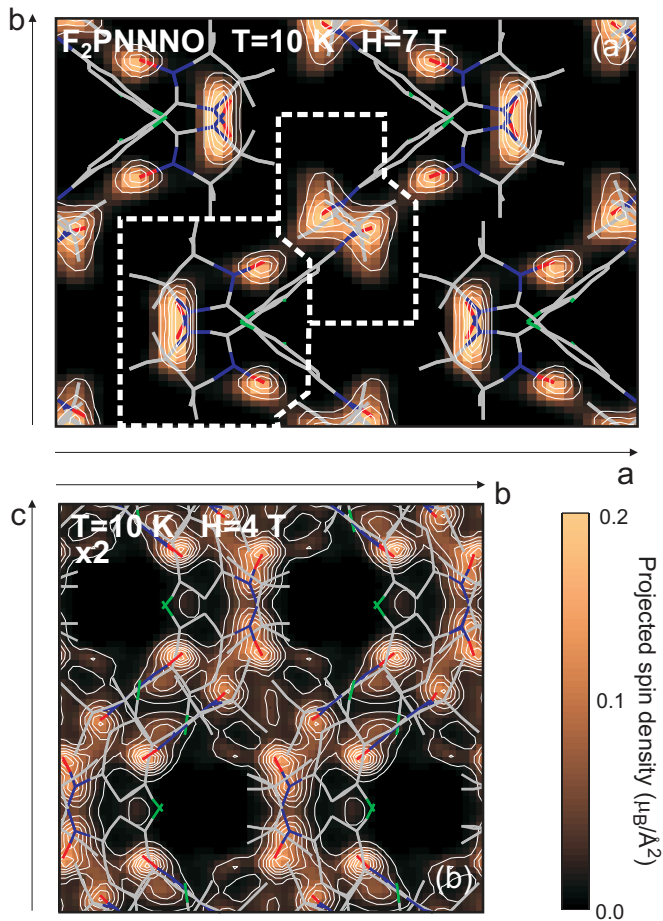}
\caption{ (Color online) Experimental spin density distribution in
a \FPN\ spin-tetramer at $T=10$~K, as reconstructed using the
Maximum Entropy method. (a) and (b) are projections onto the
$(a,b)$ and $(b,c)$ (crystallographic planes, respectively.
Overlaid are skeletal representations of \FPN\ molecules as they
are positioned within the crystallographic unit cell. Areas
outlined with thick dashed lines were used to estimate the spin
populations on the nitronyl nitroxide and {\it tert}-butyl
nitroxide groups (see text).} \label{fig3}
\end{figure}

\begin{figure*}
\includegraphics[width=15cm]{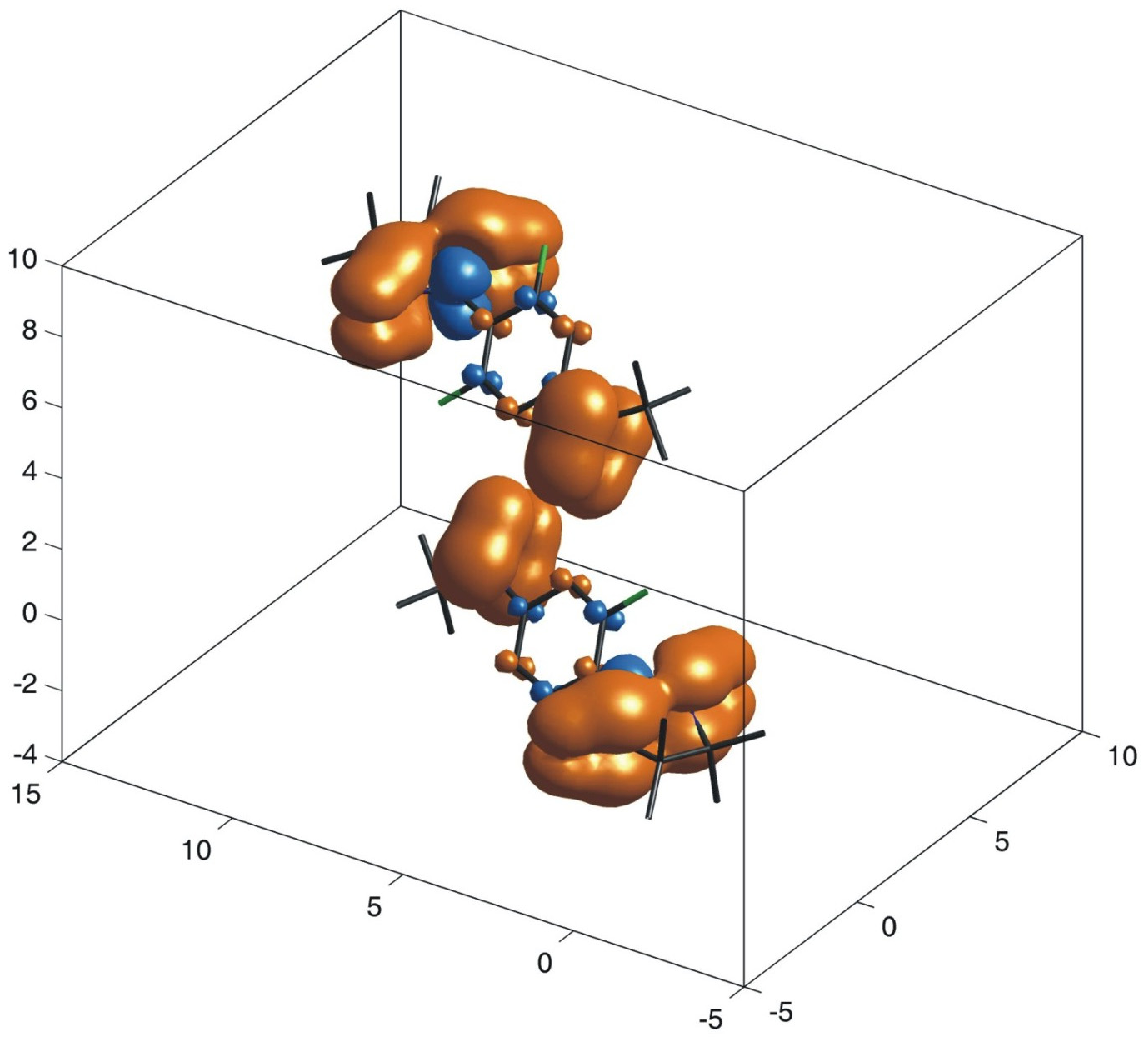}
\caption{ (Color online) Experimental spin density distribution in
a \FPN\ spin-tetramer at $H=7$~T and $T=10$~K, as reconstructed
using orbital model refinement (see text). The isosurfaces are
drawn at $1\times 10^{-3} \mu_\mathrm{B}/$\AA$^3$ (orange) and
$-1\times 10^{-3} \mu_\mathrm{B}/$\AA$^3$ (blue) levels. The axes
show cartesian coordinates in Angstroms. } \label{fig4}
\end{figure*}

\section{Results}

Inverting the Fourier transform to reconstruct the real-space
$S_z(\mathbf{r})$ function is far from straightforward, and
prompted us to apply several complimentary approaches. One such
tool was the maximum entropy (ME) method.\cite{Papoular1990} The
procedure is model-independent: it uses only the experimental
structure factors and crystal symmetry as input, and does not rely
on any additional information or assumptions. It is known to be
particularly effective at reconstructing 2D projections of
$S_z(\mathbf{r})$ onto planes close to the principal scattering
plane of the diffractometer.\cite{Papoular1995} Results of such
reconstructions for \FPN\ are shown in Fig.~2. Despite the limited
experimental spatial resolution, one immediately sees that the
spin density is primarily localized around the N and O atoms.
Already at this stage it is possible to get a crude estimate of
the redistribution effect. Integrating over the empirically chose
regions outlined in Fig.~2a, we get $R=1.22$. However, the actual
NN/tBuNO spin population imbalance is likely to be more pronounced
than suggested by ME. For all its advantages, the algorithm is
known to systematically bias the answer towards a more uniform
distribution.

An alternative reconstruction method known as atomic orbital
expansion (AOE) \cite{Gillon1989} lacks the benefit of being
model-independent, but is free of such a bias and is better
quantified. It involves refining a parameterized model for
$S_z(\mathbf{r})$ to best-fit the experimental Fourier data. Its
application to \FPN\ was founded on previous experiments
\cite{Zheludev1994,Zheludev1994-2,Davis1972,Neely1974} and
first-principle calculations \cite{Zheludev1994} for related
nitroxides. The NN spin density was described in terms of five
atomic populations. It was assumed to be concentrated in $p_z$
Slater-type atomic orbitals of the O, N and apical C atoms, the
$z$ axis chosen perpendicular to the corresponding N-O-C planes.
Two more parameters were used to quantify the spin density
delocalized over the $p_{z'}$ orbitals of the tBuNO N and O atoms,
with the $z'$ axis perpendicular to the tBuNO O-N-C plane. One
additional parameter was used for the sign-alternating spin
density induced in the phenyl ring by virtue of the spin
polarization effect.\cite{Davis1972,Neely1974} It was assumed to
be contained in $p_{z''}$ orbitals of the phenyl's C atoms ($z''$
is oriented perpendicular to the phenyl plane). The final three
parameters were the radial exponents of Slater-type atomic
orbitals for the N, O and C atom types. This model yields an
excellent least-squares fit to the data collected at $H=7$~T and
$H=4$~T, with $\chi^2=1.09$ and $\chi^2=1.20$, respectively.
Figure~2 is an isosurface representation of the resulting
3-dimensional spin density distribution in the tetramer at
$H=7$~T. Individual atomic spin populations obtained in the
refinement are listed in Table~\ref{populations}.

 \begin{table}%[H] add [H] placement to break table across pages
 \caption{\label{populations} Experimental atomic spin populations ($\mu_B$ units)
 obtained using the AOE reconstruction method. }
 \begin{ruledtabular}
 \begin{tabular}{l l l l }
 & $H=4$~T & $H=7$~T\\
 \hline
 nitronyl nitroxide\\
 N1 & 0.021(6)  &  0.047(9)\\
 O1 & 0.026(4)  &  0.044(5)\\
 N2 & 0.012(3)  &  0.057(6)\\
 O2 & 0.033(4)  &  0.031(5)\\
 C1 & -0.008(4) & -0.033(5)\\
\hline
{\it tert}-butyl nitroxide\\
 N3 & 0.025(4)  &  0.040(7)\\
 O3 & 0.031(4)  &  0.055(6)\\
\hline
 Phenyl C-atoms & $\pm$ 0.0005(5) & $\pm$0.015(5)\\
 \end{tabular}
 \end{ruledtabular}
 \end{table}

A very good measure of the AOE's reliability is its result for the
total tetramer magnetization: $m=0.48(2)\mu_\mathrm{B}$ and
$m=0.28(2)\mu_\mathrm{B}$, for $H=7$~T and $H=4$~T, respectively,
at $T=10$~K. These values are consistent with existing bulk
susceptibility data, and agree well with a thermodynamic
quantum-mechanical calculation for a single tetramer:
$m=0.59\mu_\mathrm{B}$ and $m=0.32\mu_\mathrm{B}$, respectively.
With this assurance of the validity of our approach, we can
finally obtain experimental estimates for the imbalance between
the NN and tBuNO spin populations: $R=1.53(3)$ and $R=1.51(2)$,
for $H=7$~T and $H=4$~T, based on AOE model refinements.

\section{Discussion}
A value $R>1$ signifies a reduction of {\it uniform} induced
magnetization around an {\it antiferromagnetic} bond in the
tetramer, and is to be expected. What is important though, is that
this effect is hugely magnified by quantum correlations. In a
classical magnet with similar exchange constants, all spins would
align themselves in the $(x,y)$ plane and tilt slightly in the
field ($z$) direction. It is easy to show that the resulting
imbalance in $S_z(\mathbf{r})$ would be an order of magnitude
smaller: $R_\mathrm{classical} \approx
1+|J_\mathrm{AF}|/4|J_\mathrm{F}|=1.04$.

The quantitative agreement between experiment and direct
diagonalization calculations is an important microscopic
validation of the model Heisenberg Hamiltonian that was initially
hypothesized based on bulk measurements alone.\cite{Hosokoshi1999}
In fact, our measurements can be viewed as a direct experimental
determination of $J_\mathrm{F}/J_\mathrm{AF}$ in \FPN. In
addition, these experiments help understand the microscopic
interactions within the \FPN\ molecule. In Fig.~3, note the
negative density in the vicinity of the apical carbon atom of the
NN group. This large negative spin population\cite{Davis1972}
plays a key role in the ferromagnetic intra-molecular coupling
$J_\mathrm{F}$. It is a part of a sign-alternating spin density
wave that propagates across the phenyl ring and connects the
positively populated N sites of the NN and tBuNO fragments over a
large distance. This density-wave mechanism is analogous to
Ruderman-Kittel-Kasuya-Yosida interactions in
metals.\cite{Ruderman1954,Yosida1957}

\section{Concluding remarks}
Field-induced spin distributions in quantum magnets are strongly
affected by quantum correlations. They can be directly probed by
polarized neutron diffraction and carry valuable information on
the system. A very promising avenue for future work are similar
experiments on \FPN\ conducted at low temperature in the
magnetization plateau phase $H> H_{c2}$ and in the inter-plateau
region $H_{c1}<H<H_{c2}$. In these regimes the system is expected
to be ordered in three dimensions due to inter-tetramer
interactions. How does transverse long-range order influence the
distribution of $S_z$, and is in any different from that in
effectively isolated tetramers, as studied in this work?

\acknowledgements{
 The authors thank Dr. E. Ressouche
(CEA Grenoble) for his expert assistance with the data analysis
software.  Work at ORNL was funded by the United States Department
of Energy, Office of Basic Energy Sciences- Materials Science,
under Contract No. DE-AC05-00OR22725 with UT-Battelle, LLC. This
work was supported in part by Grant-in-Aid for Scientific Research
(B) No.18350076 and on priority Areas "High Field Spin Science in
100T"(No.451) from the Ministry of Education, Culture, Sports,
Science and Technology(MEXT) of Japan.}

%=====================================================================
%=====================================================================
%=====================================================================
%=====================================================================
%=====================================================================

%\bibliography{nhaldane}

%=====================================================================
%=====================================================================
%=====================================================================
%=====================================================================
%=====================================================================

%=====================================================================
%=====================================================================
%=====================================================================
%=====================================================================
%=====================================================================

\end{document}